\documentstyle[preprint,aps,floats,psfig]{revtex}
\begin{document}
\hsize\textwidth\columnwidth\hsize\csname@twocolumnfalse\endcsname

\draft

\title{Theory of sound attenuation in glasses: The role of thermal
vibrations.}
\author{Jaroslav Fabian$^{1,2}$  and Philip B. Allen$^2$}
\address{$^1$Department of Physics, University of Maryland at College Park,
College Park, MD 20742-4111}
\address{$^2$Department of Physics and Astronomy, State University of New
York at Stony Brook, Stony Brook, New York 11794-3800}
\maketitle

\vspace{1cm}

\begin{abstract}
Sound attenuation and internal friction coefficients are calculated
for a realistic model of amorphous silicon. It is found that,
contrary to previous views, thermal vibrations can
induce sound attenuation at ultrasonic and hypersonic frequencies that
is of the same order or even larger than in crystals. The reason is
the internal-strain induced anomalously large Gr\"uneisen parameters
of the low-frequency resonant modes. 
\end{abstract}
\vspace{2em}
\tighten
\newpage
Sound attenuation in glasses is poorly understood. 
This is because many competing factors lead to
sound wave damping. Most important are thermally activated structural
relaxation, hypothetical tunneling states, topological defects, and
thermal vibrations. Sorting out different contributions for a given
temperature $T$ and sound wave frequency $\nu=\Omega/2\pi$ is a
difficult task.

Experiments show the following features:
(i)
At temperatures $T\alt 10$ K and ultrasonic frequencies (10 MHz to 1
GHz) the  sound attenuation coefficient $\Gamma(T)$  exhibits
a small, frequency-dependent peak
\cite{hunklinger76}. (ii) At higher temperatures, between 10
and 200 K, another peak in $\Gamma(T)$ develops whose center
increases only moderately when $\nu$ increases.  The peak
broadens at hypersonic frequencies
\cite{vacher81} and is not seen above 100 GHz \cite{morath97,zhu91}.
As a function of frequency  $\Gamma(\nu)\sim\nu$ at the peak
temperatures \cite{vacher81}.  (iii) At hypersonic frequencies
$\Gamma(T)$ appears to be almost independent of (or slightly
increasing with) $T$ above the peak (ii) to at least 300 K
\cite{morath97,zhu91}.  (iv) Room temperature $\Gamma(\nu)\sim\nu^2$
from at least 200 MHz \cite{vacher81}; this dependence continues for
up to about 300 GHz \cite{morath97,zhu91} and seems valid for any
temperature above the peak (ii). Finally, (v) the attenuation
coefficients for longitudinal ($\Gamma_L$) and transverse
($\Gamma_T$) waves are similar.
\cite{vacher81}.

While the low-temperature behavior (i) of $\Gamma$ is 
understood based on the interaction of sound waves with
tunneling states
\cite{hunklinger76}, features (ii) through (v) lack consistent
theoretical justification. The higher-temperature peak (ii)
shows many attributes of a thermally activated relaxational
process\cite{tielburger92}, but a calculation shows
that to fit experiment, different sets of
relaxational processes are needed for different $\nu$
\cite{vacher81}. Also the plateau region (iii) is difficult to explain
by a thermally-activated relaxation process since numerical
fits require unphysically large attempt frequencies\cite{zhu91}.
Further, thermal relaxation processes give attenuation
that increases more slowly than quadratic with increasing $\nu$,
contradicting (iv). Thermal vibrations
have been overlooked as a sound-wave damping factor on 
grounds that vibrational modes would need unreasonably large
Gr\"uneisen parameters ($\gamma \approx 200$ for vitreous silica
\cite{vacher81}) to account for the measured $\Gamma$.
Until now, however,  there has been no numerical study to
test this argument.

In this paper we examine the role thermal vibrations play
in the sound attenuation in glasses. 
We will use the term ``vibron'' to refer to any quantized vibrational
mode in a glass\cite{fabian96}.
Our analysis is restricted
to the region $\Omega\tau_{\rm \,in}\alt 1$ (the so called Akhiezer
regime\cite{maris71}), where $\tau_{\rm \,in}$ is the inelastic 
lifetime or thermal equilibration time of a thermal vibron.
We show that the unusually strong coupling (measured by Gr\"uneisen
parameters $\gamma$) between sound waves and the low-frequency
resonant modes explains the features (iii) through (v).  As for the
interpretation of (ii), our calculation shows that confusion
arises because there actually are two different peaks.  One is caused by
relaxational processes (not addressed here) and dominates below 1 GHz
and another is due to thermal vibrations and dominates at the lowest
hypersonic frequencies.  A double peak structure should be expected
at intermediate frequencies.  There is some indication for such structure
in measurements on vitreous silica\cite{vacher81}.
Our calculation is also a prediction: the existing measurements
on amorphous Si\cite{haumeder80} report $\Gamma$ at too low 
frequencies (300 MHz) to 
see contributions of thermal vibrations. But even at higher
frequencies (say, 30 GHz) one may expect traces of thermally 
activated peaks due to various defects. Recently discovered 
amorphous Si with 1 at.  \% H \cite{liu97} in which tunneling 
(and perhaps also relaxational) processes are
inhibited would be excellent to test our results.

In the Akhiezer regime a sound wave passing through
a solid can be attenuated by two processes\cite{gurevich86}. First,
if the wave is longitudinal, periodic contractions
and dilations in the solid induce a temperature wave via
thermal expansion. Energy is dissipated by heat conduction between
regions of different temperatures. Second, dissipation occurs as the
gas of vibrons tries to reach an equilibrium characterized by a local
(sound-wave induced) strain. This is the internal friction mechanism.
To establish the relative importance of the two processes, consider
order-of-magnitude formulas $\Gamma_h\approx
(\Omega^2/\rho v^3)(\kappa T \alpha^2 \rho^2 v^2/C^2)$ and
$\Gamma_i\approx(\Omega^2/\rho v^3)(CT\tau_{\rm \,in}\gamma^2)$ for
the heat conductivity and internal friction processes,
respectively\cite{gurevich86}. Here $\rho$ is density, $C$ specific
heat per unit volume, $v$ sound
velocity, $ \kappa$ thermal conductivity, and $\alpha$ is the
coefficient of thermal expansion. The ratio $\Gamma_h/\Gamma_i\approx
(\kappa\alpha^2\rho^2 v^2)/(C^3\tau_{\rm \,in}\gamma^2)$ becomes more
intuitive when putting $\alpha\approx C\gamma/B$ ($B\approx
\rho v^2$ is the bulk modulus) and $\kappa\approx CD$ where $D$ is
diffusivity.  Then $\Gamma_h/\Gamma_i\approx D/(v^2 \tau_{\rm \,in})$.
The factor $v^2\tau_{\rm \,in}$ measures the ability of vibrons to
absorb energy from a sound wave of velocity $v$. The 
difference between a glass and a crystal lies in the values
of $D$ and $\tau_{\rm \,in}$. In crystals $D\approx v^2\tau_{\rm \,in}$, that is,
energy is carried by phonon wave packets with group velocity 
$v$. The ratio $\Gamma_h/\Gamma_i$ is then of order unity. In
glasses energy is transferred by diffusion (spreading rather than
ballistic propagation of wave packets\cite{kelner98}) and $D$ is not related to
$\tau_{\rm \,in}$\cite{allen93}.  
One of the reasons the contribution to
$\Gamma_i$ of thermal vibrons was previously underestimated is that
$\tau_{\rm \,in}$ was guessed from thermal conductivity \cite{vacher81};
this gave too small $\tau_{\rm \,in}$. For amorphous Si $D\approx
10^{-6}$ ${\rm m}^2/s$\cite{allen93}, $v\approx 8\times 10^3$ m/s,
and $\tau_{\rm \,in}\approx$ $10^{-12}$s\cite{fabian96} give
$\Gamma_h/\Gamma_i\approx$ 0.02.
Since these are typical values, $\Gamma_h$ can be neglected. This is
consistent with experiment: compared with crystals glasses have
smaller $\kappa$ and yet $\Gamma$ can be larger \cite{vacher81}.

Internal friction leads to sound-wave energy attenuation
\cite{gurevich86} $
\Gamma=(\Omega^2/\rho v^3 q^2)\eta_{\alpha\beta\gamma\delta}
{q}_{\alpha}{e}_{\beta}{q}_{\gamma}{e}_{\delta}$,
where $\eta_{\alpha\beta\gamma\delta}$ is the internal
friction tensor with cartesian coordinates $\alpha...\delta$
and ${{\bf q}}$ (${{\bf e}}$) is the wave vector
(polarization) of the sound wave.
Summation over repeated indexes is assumed.
We will evaluate $\Gamma$ for both longitudinal
($L$) and transverse ($T$) sound waves with
wave vectors averaged over all directions:
\begin{eqnarray} \label{eqn:1}\Gamma_L=\frac{\Omega^2}{15\rho
v^3_L}
(\eta_{\alpha\alpha\beta\beta}+2\eta_{\alpha\beta\alpha\beta}),\\
\label{eqn:2}\Gamma_T=\frac{\Omega^2}{30\rho v^3_T}
(3\eta_{\alpha\beta\alpha\beta}-\eta_{\alpha\alpha\beta\beta}).
\end{eqnarray}
The coefficients $\eta_{\alpha\beta\gamma\delta}$ are the real part
of a complex tensor $\bar{\eta}_{\alpha\beta\gamma\delta}$ which can
be obtained by solving a kinetic equation in relaxation time
approximation \cite{maris71},
\begin{eqnarray}
\label{eqn:3}
\bar{\eta}_{\alpha\beta\gamma\delta}=
\sum_jTc_j\tau_j\frac{
\gamma_{\alpha\beta}^j\gamma_{\gamma\delta}^j-
\left(\bar{\gamma}_{\alpha\beta}
\gamma_{\gamma\delta}^j+\gamma_{\alpha\beta}^j
\bar{\gamma}_{\gamma\delta}\right)/2}
{1-i\Omega\tau_j}.
\end{eqnarray}
The summation is over all vibrational modes $j$; $c_j$ and
$\tau_j$ denote mode specific heat and relaxation time.
The Gr\"uneisen tensor $-\gamma_{\alpha\beta}^j$ is the relative
shift of mode frequency $\omega_j$ per unit strain 
$\epsilon_{\alpha\beta}$; $\bar{\gamma}$ is the mode average 
of $\gamma^j$ weighted with
$c_j/(1-i\Omega\tau_j)$. The applicability of kinetic theory to
the problem of internal friction was justified by DeVault and
coworkers \cite{devault65} who obtained $\eta$ from a microscopic
theory as an autocorrelation function of the momentum current density
operator.
Remarkably, the microscopic theory shows that the momentum current in
a solid is not monopolized by ballistically propagating
vibrational modes as in the case of the
energy current.  Nonpropagating (even localized) modes can
contribute as much as propagating ones to the momentum current. 
One consequence is that
the concept of ``minimum kinetic coefficient,'' as introduced for
electrical\cite{mott70} or heat\cite{kittel49} conductivity of
disordered systems, is not realized for internal friction.
We generalized \cite{fabian97b} DeVault's theory to include internal
strain, the atomic rearrangements in a strained solid. 
We found that internal strain affects internal friction only by 
modifying $\gamma^j$, as in the case of thermal expansion\cite{fabian97}:
$\gamma^j$ now reflects the change between
the initial mode frequency and the frequency of the mode after the
rescaling (scaling parameter $1+\epsilon$) {\it plus}
the rearranging of atomic positions (to achieve a new
equilibrium at strain $\epsilon$). 
Internal strain is very important for thermal expansion of
glasses\cite{fabian97}; we will show that it is important for
$\eta$ (and $\Gamma$) as well.

We calculate $\eta$ and $\Gamma$ for the model of
amorphous Si based on the Wooten-Winer-Weaire
atomic coordinates \cite{wooten85} and Stillinger-Weber interatomic
forces \cite{stillinger85}, with 1000 atoms arranged in a cube of 
side 27.549 \AA\, with periodic
boundary conditions. Diagonal Gr\"uneisen parameters
$\gamma_{\alpha\alpha}^j/3\equiv (\gamma_{11}^j+\gamma_{22}^j+
\gamma_{33}^j)/3 $
for this model\cite{feldman98} were given
in Ref.  \cite{fabian97}, transverse $\gamma_{\alpha\beta}$ are
calculated here. Vibrational lifetimes $\tau_j$ are extracted from
their 216-atom version values\cite{fabian96} (see also Ref.
\cite{bickham98}). The model has sound velocities $v_L=7640$ m/s
and $v_T=3670$ m/s\cite{feldman91}.

Figure \ref{fig:1} shows the calculated $\Gamma(\nu)$ for longitudinal
and transverse sound waves in amorphous Si from 10 MHz to
1 THz at 300 K.  The attenuation $\Gamma\sim \nu^2$ up to about
100 GHz, where the condition for the applicability of kinetic theory,
$\Omega\tau_{\rm \,in}\alt 1$ reaches its limit ($\tau_{\rm \,in}\approx$ 1 ps).
Our calculation is not valid beyond this point. In comparison,
the measured attenuation of longitudinal waves in vitreous silica
grows quadratically with $\nu$ up to at least 400 GHz
\cite{zhu91} suggesting that $\tau_{\rm \,in}$ in vitreous silica is
several times smaller than in amorphous Si. This is
not surprising since Si is remarkably harmonic: room
temperature heat conductivity of crystalline Si is larger by an
order of magnitude than that of quartz\cite{mason65}, and 
a similar relation may hold for the
corresponding $\tau_{\rm \,in}$ of the glassy phases.

More surprising is the comparison with crystalline Si.
Figure \ref{fig:1} shows that $\Gamma_L$ is
similar for the amorphous and crystalline cases
(measured $\Gamma$ for vitreous silica
is several times larger than for quartz\cite{vacher81}).
One would  naively
expect the sound attenuation in a glass to be much smaller than in
the corresponding crystal since, owing to a distribution of bond
lengths and bond angles, anharmonicity of the glass is higher (and
$\tau_{\rm \,in}$ smaller). The same interatomic potential, for example,
yields $\tau_{\rm \,in}$ for high-frequency phonons
in crystalline Si at 300 K about five times larger than in
amorphous Si\cite{fabian96}. The reason why $\Gamma$ in glasses
can be of the same order or even higher than in crystals is the
internal-strain induced anomalously large Gr\"uneisen parameters of
the resonant modes\cite{fabian97} (see also Fig.\ref{fig:3}).
(Resonant modes are low-frequency extended modes whose amplitude is 
unusually large at a small, typically undercoordinated region
\cite{fabian97,biswas88}.)
Atomic rearrangements caused by internal
strain are largest in the same regions  of undercoordination
where the resonant modes have largest amplitude
\cite{fabian97}. This leads to high sensitivity 
(measured by
$\gamma$) of the frequencies of these modes to strain.  If the
internal strain is neglected, the sound attenuation is an order of
magnitude smaller, as seen in Fig.  \ref{fig:1}.  (Since the resonant
modes have low frequencies, their $\tau_j$ is longer than an average
$\tau_{\rm \,in}$; this adds even more weight to these modes.) Fewer than
one percent of the modes are capable of increasing $\Gamma$ by a
decade!  We believe the measured $\Gamma$ for vitreous silica is also
caused by the strong coupling of sound waves and resonant modes.
Vitreous silica is a much more open structure than amorphous Si
so the number of resonant modes should be higher, bringing $\Gamma$
above the crystalline value.

Another interesting feature in Fig.\ref{fig:1} is the relative
attenuation strength for longitudinal and transverse sound waves.
While our model of amorphous Si gives $\Gamma_L/\Gamma_T\approx
1/3$ at 300 K, the measured ratio for crystalline Si is reversed:
$\Gamma_L/\Gamma_T\approx 3$ \cite{mason65}. This again shows
how differently is sound attenuated in glasses and in crystals.
The ratio $\Gamma_L/\Gamma_T$ can be written
as $(v_T/v_L)^3(\gamma_L^2/\gamma_T^2)$, where $\gamma_L$ and
$\gamma_T$ are effective Gr\"uneisen parameters. 
A crude way to
estimate $\gamma_L^2$ and $\gamma_T^2$, suggested by Eqs. \ref{eqn:1}
and \ref{eqn:2} is to take mode averages  of
$(\gamma_{\alpha\alpha}^j \gamma_{\beta\beta}^j+
2\gamma_{\alpha\beta}^j\gamma_{\alpha\beta}^j)/15$ and
$(3\gamma_{\alpha\beta}^j
\gamma_{\alpha\beta}^j-
\gamma_{\alpha\alpha}^j\gamma_{\beta\beta}^j)/30$.
Our model gives $\gamma_L^2\approx 3$ and $\gamma_T^2\approx 1$.
The ratio $\Gamma_L/\Gamma_T$ is then about 1:3, in accord with the
full calculation. Assuming the same ratio
$\gamma_L^2/\gamma_T^2\approx 3$  for vitreous silica
($v_L=5800$ m/s and $v_T=3800$ m/s), transverse
and longitudinal waves are attenuated about equally. This is observed
in experiment\cite{vacher81}. The explanation of the measured
$\Gamma_L/\Gamma_T$ in crystalline Si can be found in Ref.
\cite{mason65}.

In Fig. \ref{fig:2} we plot $\Gamma(T)$ for different $\nu$.
A remarkable feature is a peak at about 20 K
at 1MHz and below. As $\nu$ increases, the peak shifts towards
higher $T$ and vanishes above 4-5 GHz. Two factors cause the
peak. (a) The sum $\sum_j c_j(\gamma^j)^2$
saturates at much lower temperatures (about 50 K) than
the model Debye temperature $T_D\approx 450$ K\cite{feldman91}. 
This is because
the relevant $j$ are resonant modes with small
frequencies. (b) For low-frequency modes $T\tau_j$, after
increasing linearly develops a peak, before going constant [much like
$\Gamma(T)$ itself]. As the temperature dependence of $\Gamma$
follows $\sum_jc_j(\gamma^j)^2\ T\tau_j$, the peak appears.
At large $\nu$ the peak vanishes because of the factor
$1/(1+\Omega^2\tau^2)$ in Eq. \ref{eqn:3}. At $T$ above 100 K
$\Gamma(T)$ is nearly constant, as observed in experiment as
 a plateau (iii). This again follows from (a) and (b).

We are not aware of any experiment with which we could compare
our calculations. The measurement of $\Gamma(T)$ of sputtered
amorphous Si films reported in Ref. \cite{haumeder80}, for example, 
was performed at 300 MHz. This is too low to see any contributions
from thermal vibrations. The whole temperature spectrum is dominated
by a single peak of the type (ii), except at very low temperatures.
This peak is expected to increase linearly with $\nu$, until
thermal vibrations become relevant (roughly at 10 GHz), causing 
a plateau (iv) that increases as $\nu^2$ at higher
frequencies. Even
at smaller frequencies one may see some 
vibrational contribution to $\Gamma(T)$ at large enough $T$, since  
the thermally activated peak decreases
as $1/T$ at large $T$. 

Anomalous low $T$ thermal expansion already suggested\cite{white75} 
very large $\gamma$ values for low $\omega$ modes. Our large $\gamma$ values
\cite{fabian97}
agree nicely with trends in $\alpha(T)$. 
Like thermal expansion, $\Gamma$ should
be strongly sample and model dependent.  
There is evidence\cite{feldman99} that our 
highly homogeneous model of amorphous Si becomes free of resonant 
modes when the number of atoms grows to infinity. That means
an infinite model would predict $\Gamma$ about a decade smaller 
than calculated here. Amorphous silicon, however, can be prepared 
only in thin films where voids and other inhomogeneities are 
unavoidable. Voids loosen the strict 
requirements of a tetrahedral random network (for example by 
introducing free boundary conditions). Then, as in our 
finite models, regions of undercoordinated atoms will allow 
the formation of resonant modes. 
While this issue for amorphous silicon will be ultimately 
settled by experiment, our calculation combined with the existing data 
on vitreous silica strongly suggests the reality of resonant modes.

Our final note concerns the mode dependence of transverse Gr\"uneisen
parameters like $\gamma_{12}$. Similarly to
volumetric $\gamma_{\alpha\alpha}/3$\cite{fabian97}, transverse
$\gamma_{12}$ in Fig. \ref{fig:3} ($\gamma_{13}$ and $\gamma_{23}$ 
look the same) is unusually large for resonant
modes and have scattered values for high-frequency
localized modes. (More resonant modes have
$\gamma_{12}$ negative than positive which suggests that 
resonant modes are trapped at  highly anisotropic undercoordinated
regions whose sizes change under shear\cite{fabian97}.) The $15-70$
meV vibrons (diffusons\cite{fabian96}) have $\gamma_{12}\approx 0$ 
(average magnitude 0.02), while the corresponding 
$\gamma_{\alpha\alpha}/3$ are of order unity\cite{fabian97}.  
Such small values (zeros in an infinite
model) are characteristic for diffusons, 
which are extended modes whose polarization directions (atomic
displacements) point, in general, at random. There remains only a
short-range correlation between polarization directions which determines
the diffuson's frequency $\omega_d$. If a shear, say, $\epsilon_{12}$
is applied, $\omega_d$ changes to $\omega_d'(\epsilon_{12})$. Since
long-range order in the diffuson polarization is absent,
$\omega_d'(\epsilon_{12})\approx\omega_d'(-\epsilon_{12})$, and
$\gamma_{12}$ which is a linear coefficient in the expansion of
$\omega_d'$ in $\epsilon_{12}$ must vanish.

We thank J. L. Feldman for helpful discussions. The work
was supported by NSF Grant No. DMR 9725037. J. F. acknowledges
also support from the U.S. ONR.

\newpage

\begin{figure}
\caption{Log-log plot of the sound attenuation
$\Gamma$(cm$^{-1}$) at 300 K, as a function of sound-wave frequency
$\nu$(Hz). Calculated $\Gamma$ are represented by lines (IS--internal
strain), experimental data by symbols. L (T) stand for longitudinal
(transverse) sound waves.
}
\label{fig:1}
\end{figure}

\begin{figure}
\caption{Calculated sound attenuation $\Gamma(T)$ for amorphous
Si at different frequencies. The thin dashed lines are for
longitudinal waves with the labeled $\nu$ in GHz. Plotted are
rescaled values $\Gamma/\nu^2$ for $\nu$ measured in GHz.
}
\label{fig:2}
\end{figure}

\begin{figure}
\caption{Calculated transverse Gr\"uneisen parameters $\gamma_{12}$
for amorphous Si as
a function of vibron frequency. Above the vertical line
($\approx$71 meV) modes are localized.
}
\label{fig:3}
\end{figure}


\begin{references}
\bibitem{hunklinger76} S. Hunklinger and W. Arnold, in
	           {\it Physical Acoustic}, edited by W. P.
	           Mason and R. N. Thurston (Academic, New
	           York, 1976), p. 155.
\bibitem{vacher81}   R. Vacher, J. Pelous, F. Plicque, and A.
	         Zarembowitch, J. Non-Cryst. Solids {\bf 45},
	         397 (1981).
\bibitem{morath97} C. J. Morath and H. J. Maris, Phys. Rev. B {\bf
	      54}, 203 (1996).
\bibitem{zhu91}  T. C. Zhu, H. J. Maris, and J. Tauc, Phys. Rev. B
	     {\bf 44}, 4281 (1991); H. N. Lin, R. J. Stoner,
	     H. J. Maris, and J. Tauc, J. Appl. Phys. {\bf 69},
	     3816 (1991).
\bibitem{tielburger92} D. Tielb\"urger, R. Merz, R. Ehrenfels, and S.
                   Hunklinger, Phys. Rev. B {\bf 45}, 2750 (1992).
\bibitem{fabian96} J. Fabian and P.  B.  Allen, Phys. Rev.  Lett.
                   {\bf 77}, 3839 (1996)
\bibitem{maris71} H. J. Maris, in {\it Physical Acoustics}, edited by W.
                  P. Mason and  R. N. Thurston (Academic, New York,
	      1971), Vol.  VIII.
\bibitem{haumeder80} M. Von Haumeder, U. Strom, and S. Hunklinger, 
                    Phys. Rev. Lett. {\bf 44}, 84 (1980). 
\bibitem{liu97} X. Liu, B. E. White, R. O. Pohl, E. Iwanizcko,
                K. M. Jones, A. H. Mahan, B. N.
	    Nelson, R. S.  Crandall, and S. Veprek, Phys. Rev.
	    Lett. {\bf 78}, 4418 (1997).
\bibitem{gurevich86} V. L. Gurevich, {\it Transport in Phonon
                     Systems} (North-Holland, Amsterdam, 1986).
\bibitem{kelner98}  P. B. Allen and J. Kelner, Am. J. Phys. {\bf 66},
	        497 (1998).
\bibitem{allen93} P. B. Allen and J. L. Feldman, Phys. Rev. B {\bf
                  48}, 12 581 (1993); J. L. Feldman, M. D. Kluge,
 	      P. B. Allen, and F. Wooten, Phys. Rev. B {\bf 48},
 	      12 589 (1993).
\bibitem{devault65} G. P. DeVault and J. A. McLennan, Phys.  Rev.
                    {\bf 138}, A856 (1965); G.  P.  DeVault, Phys.
                    Rev. {\bf 149}, 624 (1966); {\bf 155}, 875
                    (1967); G.  P.  DeVault and R.  J.  Hardy,
                    Phys. Rev. {\bf 155}, 869 (1967).
\bibitem{mott70}  N. F. Mott, Philos. Mag. {\bf 22}, 7 (1970).
\bibitem{kittel49} C. Kittel, Phys. Rev. {\bf 75}, 972 (1949).
\bibitem{fabian97b} J. Fabian, Ph.D. Thesis, SUNY, Stony Brook, NY,
                    1997.
\bibitem{fabian97} J.  Fabian and P.  B.  Allen, Phys.  Rev.  Lett.
                   {\bf 79}, 1885 (1997).
\bibitem{wooten85} F.  Wooten, K.  Winer, and D. Weaire, Phys. Rev.
                       Lett. {\bf 54}, 1392 (1985).
\bibitem{stillinger85} F. H.  Stillinger and T. A.  Weber, Phys.
	           Rev.  B {\bf 31}, 5262 (1985).
\bibitem{feldman98}  Ref. \cite{fabian97} used a slightly
	         different version with a small
	         amount of residual stress.  The main results,
	         however, remain unchanged.  J. L.  Feldman,
	         private communication.
\bibitem{bickham98} S. R. Bickham and J. L. Feldman, Phys. Rev. B
	        {\bf 57}, 12 234 (1998); Philos. Mag. B {\bf 77},
	        513 (1998).
\bibitem{feldman91} J. L. Feldman, J. Q. Broughton, and F. Wooten,
	        Phys. Rev. B {\bf 43}, 2152 (1991).
\bibitem{mason65} W. P. Mason, in {\it Physical Acoustics}, edited by
              W.  P. Mason (Academic, New
              York, 1965), Vol.  III.
\bibitem{biswas88} R.  Biswas, A.  M.  Bouchard, W. A.  Kamikatahara,
                   G.  S. Grest, and C.  M.  Soukolis, Phys. Rev.
                   Lett.  {\bf 60}, 2280 (1988); H.  R.  Schober
	       and B.  Liard, Phys.  Rev.  B
                    {\bf 44}, 6746 (1991); H.  R.  Schober and C.
                    Oligschleger, Phys.  Rev. B {\bf 53}, 11 469
	        (1996).
\bibitem{white75} G. K. White, Phys. Rev. Lett. {\bf 34}, 204 (1975).
\bibitem{feldman99}  J. L. Feldman, P. B. Allen, S. R. Bickham
	         (unpublished).


\end{references}
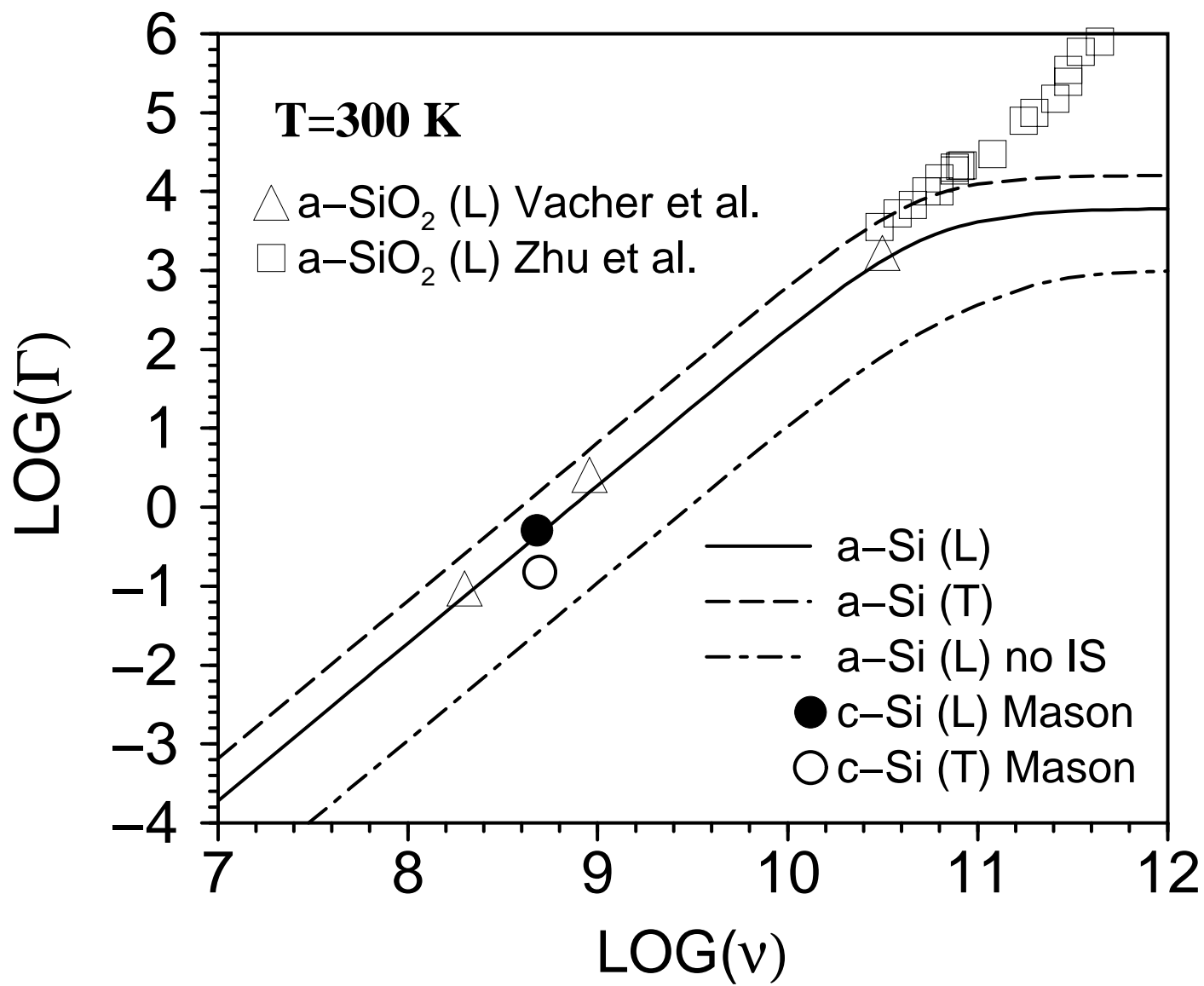
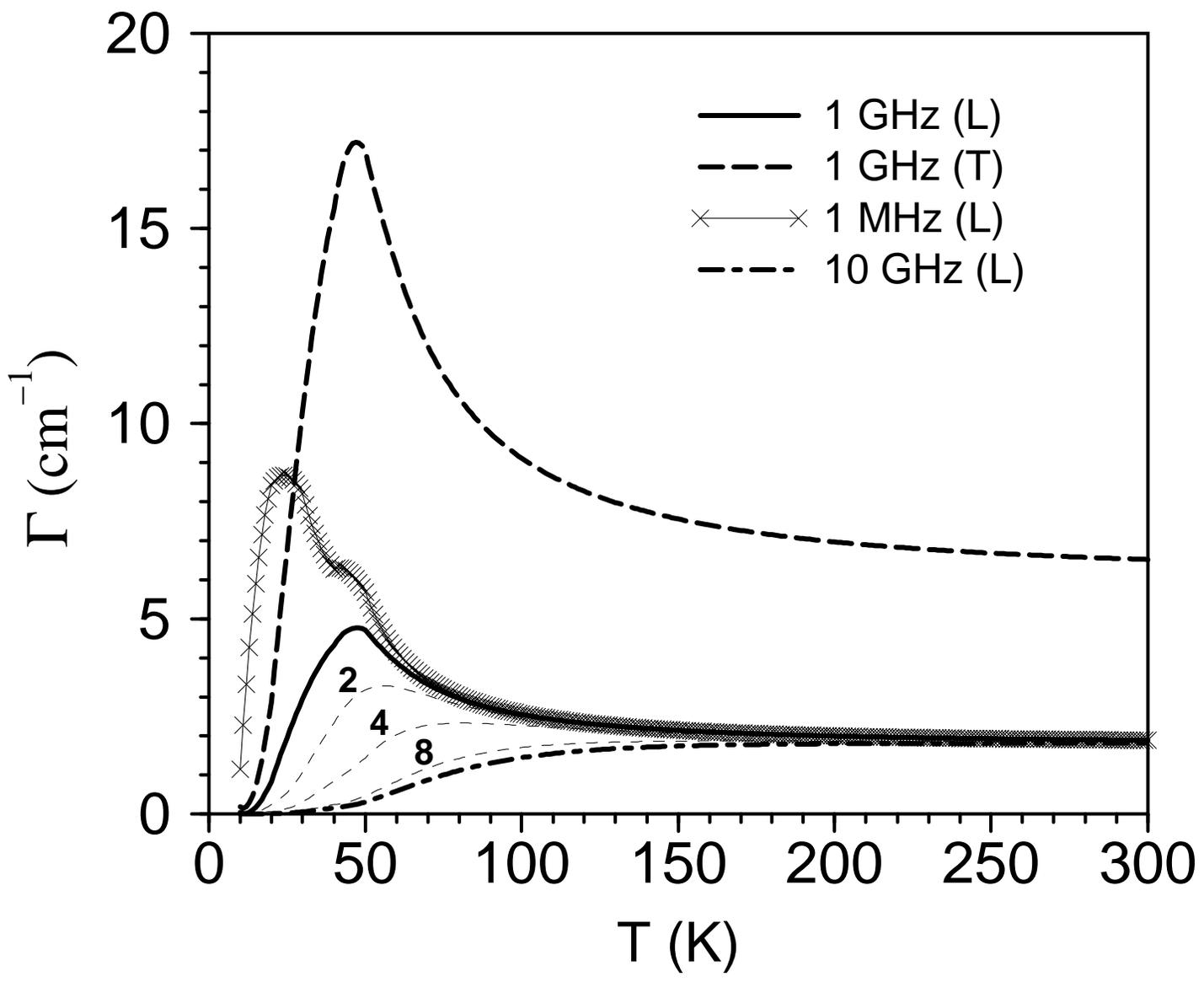
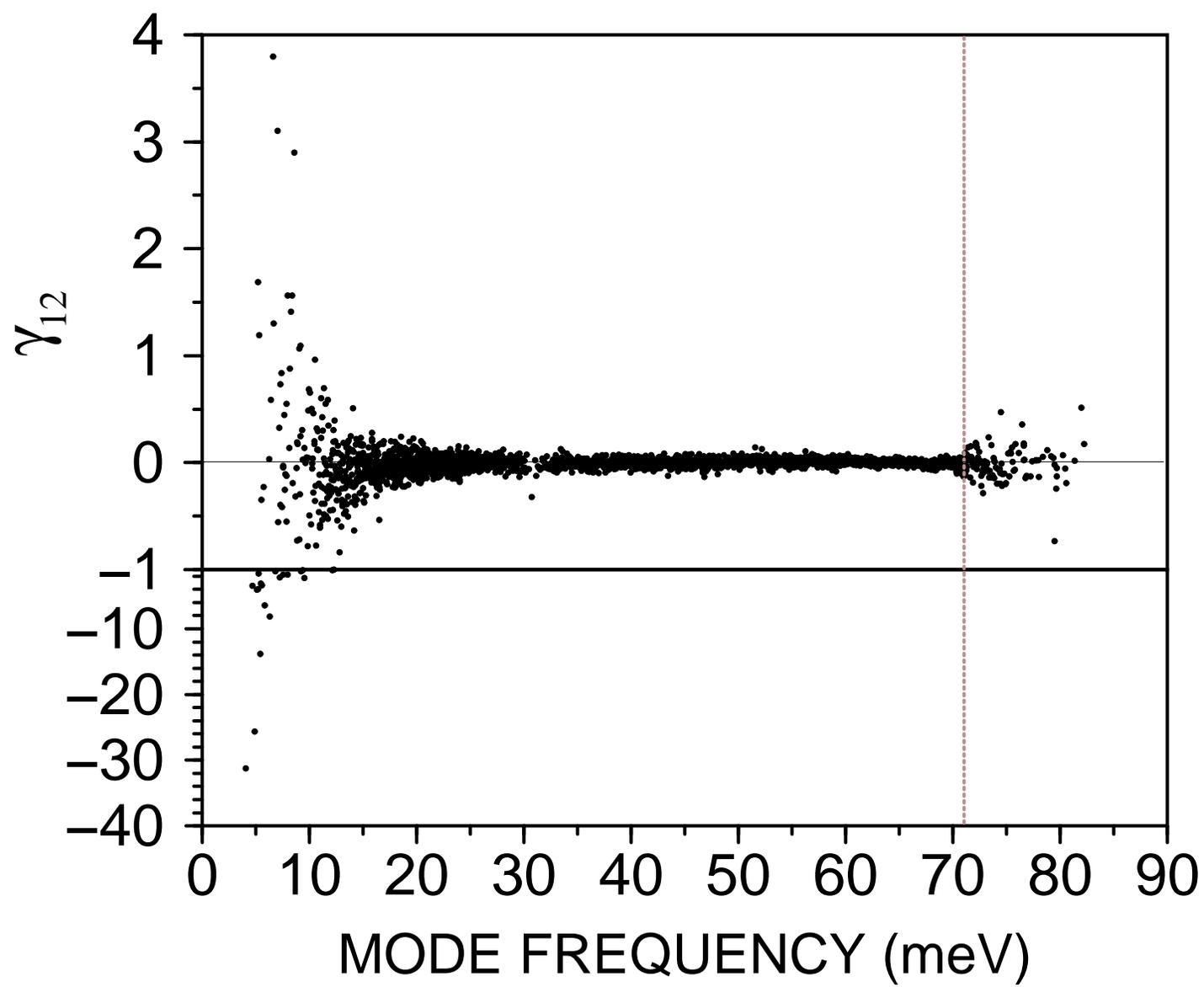
\end{document}